\documentclass[aps,preprint,showkeys]{revtex4-2}
\usepackage{graphicx}
\usepackage{epstopdf}
\usepackage{amsmath}
\usepackage{makeidx}
\usepackage{amsfonts}
\usepackage{amssymb}
\usepackage{mathtools}
\usepackage{xcolor}
\usepackage{tabularx}

\begin{document}

\title{Optimal transport by a Lagrangian dynamics of population distribution}
\date{\today}

\author{Babak Benam}
\email{babak.benam@gmail.com}
\affiliation{Department of Physics, College of Science, Shiraz University, Shiraz 71454, Iran}

\author{Abolfazl Ramezanpour}
\email{aramezanpour@gmail.com}
\affiliation{Department of Physics, College of Science, Shiraz University, Shiraz 71454, Iran}
\affiliation{Leiden Academic Centre for Drug Research, Faculty of Mathematics and Natural Sciences, Leiden University, PO Box 9500-2300 RA Leiden, The Netherlands}

\begin{abstract}
Human mobility, enabled by diverse transportation modes, is fundamental to urban functionality.	Studying these movements across scales—from microscopic to macroscopic—yields valuable insights into urban dynamics. Local adaptation and (self-)organization in such systems are expected to result in dynamical behaviors that are represented by stationary trajectories of an appropriate effective action. In this study we develop a Lagrangian dynamical model for movement processes, using local population functions as the coordinate variables. An efficient gradient descent algorithm is introduced to estimate the optimal Lagrangian parameters minimizing a local error function of the dynamical process. We show that even a quadratic Lagrangian, incorporating dissipation, effectively captures the dynamics of synthetic and empirical movement data. The inferred models reveal that inertia and dissipation are of comparable importance, while interactions and randomness in the movements induce significant qualitative changes in model parameters. Our results provide an interpretable and generative model for human mobility, with potential applications in movement prediction.
\end{abstract}

%\keywords{}

\maketitle

\section{Introduction}\label{S0}

Human mobility is important for access to jobs, services, and social interactions in city. Movements happen at different scales, from daily trips to regional and large-scale migration, all of which together influence the form and development of cities \cite{Barbosa2018,Du2025}. Inefficient mobilities waste time and resources, reduce productivity, and make access to opportunities unequal \cite{Dorostkar2023,Chen2025}. Movement processes are important for urban resilience; the ability of cities to respond to disasters depends on how populations can redistribute quickly and safely. Beyond its practical impacts, mobility reflects complex dynamical behaviors that emerge from countless local decisions and interactions, highlighting the need for theoretical frameworks capable of capturing such emergent patterns \cite{Song2010,Simini-nat-2012,Xu2021,Xu2023,Tan2025}.

Large-scale datasets have improved progress in human mobility research by detailed studies of travel behavior in large populations. They make it possible to rebuild trajectories at large scales and to test modeling approaches for reproducing flows and helping with planning \cite{Gallotti2024}. The approaches range from individual-level models which use for instance the exploration and return strategy and its generalizations \cite{Pappa2015,Jiang2016,Schlap2021}, to population-level models such as gravity, radiation, and intervening opportunities \cite{Yan-roys-2014,Ren2014,Kang-plos-2015,Qi-plos-2023}. These models rely on diverse data sources including GPS, mobile phone records, census surveys, and social media, and have been applied to tasks ranging from commuting-flow and traffic prediction to epidemic modeling and migration analysis. 

There are also new approaches, such as deep generative and foundation models that couple activities with locations and integrate different data sources to learn universal mobility dynamics \cite{Li2023,Liao2024,Ugurel2024,Ma2025}. Even though these models are effective in generating realistic and transferable patterns, their inner representations are mostly statistical and not easy to understand. The latent structures learned by these generative systems rarely map onto interpretable forces, constraints, or dynamical principles, which makes it hard to gain explanatory insights or to connect them with theories of human behavior and urban dynamics.

Traditional models of human mobility, such as the gravity and radiation frameworks, have offered valuable insights into large-scale transport patterns, yet their very strength lies in simplifications. On the other hand, detailed microscopic and agent-based  simulations can capture fine-grained individual decisions, yet they are computationally expensive and often too complex to yield transparent analytical insights. There are also mesocopic models that try to work with a coarse-grained mobility field \cite{Mazzoli-nc-2019,Zhong-srep-2024}. For instance, the authors in Ref. \cite{Zakharov2025} use Lagrangian mechanics for identification of migration dynamics. A Lagrangian neural network then can be used to represent an observed dynamics that is expected to follow the Euler-Lagrange equations \cite{Lutter2019,Cranmer2020}. Such studies are helpful in providing an optimization formulation of the system dynamics in terms of an action functional.

In this study, we introduce a Lagrangian framework for modeling time evolution of population distributions on networks, in which inertia, potential, and dissipation appear as explicit and interpretable components. We consider a harmonic potential with its minimum at the target distribution, which creates a restoring force pushing the system toward the desired state. The interaction matrix encodes local correlations and neighborhood effects, while a damping function represents congestion-like dissipation in the equations of motion. This approach goes beyond statistical models and provides interpretable parameters with direct operational meaning. We use this framework to address the inverse problem of reconstructing a mobility pattern by estimating the Lagrangian parameters. To this end, we develop a dynamic gradient descent algorithm that updates the parameters at each time step to minimize a local error function. The method is used to connect empirical data with interpretable dynamical functions and provides a scalable way to learn Lagrangian models from dynamical data. We empirically validate our approach on both synthetic and empirical datasets from Japan, Finland, and Madrid. The inferred parameters show different dynamical patterns in each city. We also propose indicators such as flow, dissipation, and characteristic times to show the roles of inertia, friction, and interaction in mobility. Finally, we study the model responses or dynamical susceptibilities, which quantify how small perturbations to initial conditions spread through the system. These susceptibilities highlight sensitive regions of the network and offer useful guidance for targeted interventions, scenario design, and stress testing. 

The paper is organized as follows. Section \ref{S1} present the synthetic dynamics and the empirical dynamical data we use in this study. Sections \ref{S2} and \ref{S3} introduce the Lagrangian model and the inverse problem of inferring the model parameters. The results are reported in Sec. \ref{S4} and the concluding remarks are given in Sec. \ref{S5}.

\section{The reference dynamics}\label{S1}
Here, we briefly describe the dynamical data that will be used to model by a Lagrangian dynamics in the next sections.

\subsection{The synthetic dynamics}\label{S11}
We consider a two-dimensional square lattice of linear size $L$ with $N=L\times L$ nodes indexed by $a=1,\dots,N$ or coordinates $x_a,y_a \in [0,L-1]$. The set of neighbors of node $a$ is denoted by $\partial a$ with node degree $k_a=|\partial a|$. We use the growth model of Ref. \cite{Stanley-natc-2017} to generate an initial population distribution $M_a(0)$, which closely resembles the empirical distributions (see Appendix \ref{app-A}). 

As a reference dynamics, we consider a movement process of $T$ time steps where each agent (driver) moves toward a single destination $D$ which here is the center of network at $x_D=y_D=L/2$. The movement process starts with the initial distribution of drivers $M_a(0)$ which is obtained by the above growth model. We use the dynamical model of Ref. \cite{Saad-prr-2020} to move the agents according to the distances of neighboring sites to the destination. More precisely, the probability of choosing a neighbor $b$ of site $a$ is 
\begin{align}
	p_{a\to b}=\frac{e^{-\alpha (D_b-D_a)}}{ \sum_{c\in \partial a} e^{-\alpha (D_c-D_a)}},
\end{align}
where $D_a=|x_D-x_a|+|y_D-y_a|$ is the Manhattan distance of node $a$ from the destination $D$. The parameter $\alpha\ge 0$ controls the degree of closeness to the destination. 

The waiting time $\Delta$ that the driver spends in link $(ab)$ is drawn from a Poisson distribution 
\begin{align}
P_{ab}(\Delta|\rho_{ab}(n))= e^{-\tau_{ab}}\frac{\tau_{ab}^{\Delta}}{\Delta!}.
\end{align}
The mean value $\tau_{ab}=L+h\rho_{ab}(n)$ depends on the average load $\rho_{ab}(n)$ of link $(ab)$ at time step $n$. The parameter $h\ge 0$ controls the strength of interactions in this system. One can consider the effect of population also in the probability $p_{a\to b}$ of choosing a neighbor, but to reduce the number of parameters we consider this effect only in the waiting times. Following less populated links is expected to reduce the strength of effective interactions which is modeled by parameter $h$.

The average flux of drivers which exit link $(ab)$ at time step $n$ and arrive at site $b$ is denoted by $f_{ab}(n)$. The center is a sink receiving only incoming fluxes. The initial values at time step $n=0$ are given by $f_{ab}(0)=0$ and $\rho_{ab}(0)=(1-\delta_{a,D})M_a(0)/k_a$, where $\delta_{a,b}=1$ if $a=b$, otherwise it is zero. The local population at each time step is given by $M_a(n)=\sum_{b\in \partial a}\rho_{ab}(n)$ when $a\ne D$. For the destination we have $M_D(n)=M_D(n-1)+\sum_{a\in \partial D}f_{aD}(n)$. The dynamical equations governing the loads and fluxes are given in Appendix \ref{app-A}.

\subsection{The empirical dynamics}\label{S12}
Three sets of human mobility data (Appendix \ref{app-B}) are used in this study: (i) Data set from Japan \cite{data-Japan} (ii) Data set from Finland \cite{data-Finland} (iii) Data set from Madrid \cite{data-Madrid}.

We consider population dynamics of a single day from the above data sets. The data are represented as time evolution of a coarse-grained population distribution $m_a(n)=M_a(n)/M$ for $T=48$ (Japan), and $T=24$ (Finland, Madrid) time steps on a two-dimensional square lattice of size $L=10,20$. The total population in each case is $M=90120$ (Japan), and $M=10^4$ (Finland, Madrid). For Japan we also construct an average dynamics of $T=48$ time steps by averaging over $75$ working days with average population $M=95301$. The data we use in this study are given in \cite{github-link}.

Note that working with a large number of grid points makes the local populations very small and results in relatively large variations with time in the local populations. Here we are interested to model a mesoscopic dynamics with smooth variations of population in space and time. Therefore we will mostly focus on the smaller lattice size $L=10$. Moreover, the three sets of raw data are given in different spatial scales but we will use the same lattice size ($L$) with the same number of model parameters to describe their dynamics. The data from Japan and Finland have nearly the same resolution. The data from Madrid have a smaller resolution but it is useful to see how these differences affect the model (Lagrangian) parameters.

\section{The model dynamics}\label{S2}
In this section we introduce the Lagrangian dynamics which is used to model the synthetic and empirical data. 

Consider a time-dependent population distribution $\mathbf{m}(t)=\{m_a(t):a=1,\dots,N\}$ on a two-dimensional square lattice of linear size $L$ with $N=L\times L$ nodes. The population density $m_a(t)=M_a(t)/M$ is related to local population $M_a(t)$ and total population $M$. The total population $M=\sum_a M_a(t)$ does not change with time $t\in (0,1)$. Let us write $m_a(t)=f_a(\mathbf{q}(t))$ and work with variables $q_a(t)\in \mathbb{R}$. Here we take the softmax function $f_a(\mathbf{q}(t))=\exp(q_a(t))/(\sum_b\exp(q_b(t)))$, but in general it could be any non-negative and normalized function.

Now we define the following Lagrangian
\begin{align}
\mathcal{L}[\mathbf{q}(t),\dot{\mathbf{q}}(t), t] = \frac{1}{2}\sum_{a,b} \dot{q_a}(t)I_{ab}\dot{q_b}(t) - V[\mathbf{q}(t),t],
\end{align}
where $I_{ab}$ represents a symmetric inertia matrix. For simplicity, we assume that the inertia matrix is diagonal $I_{ab}=I_{aa}\delta_{a,b}$ and independent of time.  It means that we ignore the couplings between the neighboring velocities which could be relevant in a complex dynamical process. This however reduces the model complexity and the risk of overfitting in simple but noisy processes. 
For the potential we consider a harmonic one with a unique minimum at $\boldsymbol\mu(t)$. The aim is to go from the initial population distribution $M_a(0)\to M_a(1)$, so we take $\mu_a(t)=\ln M_a(1)$. The potential then is given by
\begin{align}
V[\mathbf{q}(t), t] = \frac{1}{2}\sum_{a,b} (q_a(t)-\mu_a(t))\Lambda_{ab}(t)(q_b(t)-\mu_b(t)),
\end{align}
where the symmetric matrix $\Lambda$ determines the strength of interactions and correlations between the variables. In other word, the interaction parameters $\Lambda_{ab}$ says how much deviation $(q_b-\mu_b)$ at site $b$ affects the time variation of velocity at site $a$. The inverse of interaction matrix $\Lambda^{-1}$ is also related to the covariance of local populations in a Gaussian dynamical process.

The Lagrange equations in presence of dissipative forces are
\begin{align}
\frac{d}{dt}(\frac{\partial \mathcal{L}}{\partial \dot q_a(t)}) - \frac{\partial \mathcal{L}}{\partial q_a(t)} = -\sum_{b}\Gamma_{ab}\dot q_b(t),\hskip1cm \forall a.
\end{align}
The matrix $\Gamma$ controls the rate of dissipation in the system; $\Gamma_{ab}$ says how much velocity $\dot q_b$ at site $b$ affects the time variation of velocity at site $a$. The model parameters $\{\Lambda_{ab}, \Gamma_{ab}\}$ do in general depend on time. Moreover, we assume that the matrix elements $\Lambda_{ab}$ and $\Gamma_{ab}$ are nonzero only for neighboring sites $(ab)$ and $a=b$.
 
The equations of motion then read as follows
\begin{align} 
I_{aa}\frac{d}{dt} \dot q_a(t)=
- \sum_{b \in \{a,\partial a\}}\Gamma_{ab}\dot q_b(t)
- \sum_{b \in \{a,\partial a\}}\Lambda_{ab}(q_b(t)-\mu_b(t)).
\end{align}
To simplify the notation in the following we work with asymmetric matrices $\Gamma_{ab}/I_{aa}\to \Gamma_{ab}$ and $\Lambda_{ab}/I_{aa}\to \Lambda_{ab}$.  We use the directional and time-dependent interaction and dissipation matrices to capture population heterogeneity and geographical effects in the dynamics of local populations. The equations are solved with the following initial conditions 
\begin{align} 
q_a(0) &=\ln M_a(0),\\
\dot q_a(0) &=\frac{\dot M_a(0)}{M_a(0)},
\end{align}
for $t \in (0,1)$. The population density at any time is given by $m_a(t)=e^{q_a(t)}/(\sum_b e^{q_b(t)})$.

\section{The inverse problem}\label{S3}
In this section we use the above Lagrangian model to solve the inverse problem of finding the model that best describes a reference dynamics.
 
Suppose that we are given a reference population dynamics $\tilde{q}_a(t)=\ln \tilde{M}_a(t)$ from the synthetic or empirical data.
We look for an optimal set of parameters $\boldsymbol\theta=\{\Lambda_{ab}, \Gamma_{ab}\}$ in the Lagrangian dynamics of population $q_a(t)=\ln M_a(t)$ which minimizes the deviation from the reference dynamics:
\begin{align}
E=\frac{1}{2N}\sum_a \int dt (q_a(t)-\tilde{q}_a(t))^2.
\end{align}

\begin{figure}
	\includegraphics[width=12cm]{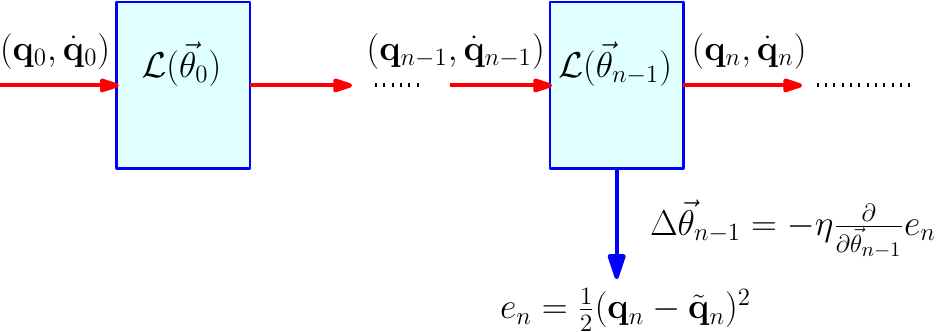} 
	\caption{Learning a Lagrangian dynamics by a local gradient descent algorithm. The model parameters $\vec{\boldsymbol\theta}_{n-1}=\{\Lambda_{ab}(n-1), \Gamma_{ab}(n-1)\}$ at time step $n$ are modified to reduce the local error function $e_n$. The model and reference dynamics are given by $\mathbf{q}_n$ and $\tilde{\mathbf{q}}_n$, respectively.}\label{fig0}
\end{figure}

First we use a discrete representation of the dynamics working with $\{q_a(n),\dot q_a(n):n=0,\dots,T\}$, where $t_n=n\Delta t=n/T$. We use the Euler-Cromer method to find an approximate solution to the Lagrange equations
\begin{align}\label{EC1} 
\dot q_a(n+1) &=\dot q_a(n)- \sum_{b \in \{a,\partial a\}}\Gamma_{ab}(n)\dot q_b(n) \Delta t
- \sum_{b \in \{a,\partial a\}}\Lambda_{ab}(n)(q_b(n)-\mu_b(n))\Delta t,\\
q_a(n+1) &=q_a(n)+\dot q_a(n+1)\Delta t.\label{EC2}
\end{align}
Second we define local error functions $e_n$ at different time steps $n$,
\begin{align}
e_n=\frac{1}{2}\sum_a (q_a(n)-\tilde{q}_a(n))^2.
\end{align}
The parameters $\Gamma_{ab}(n), \Lambda_{ab}(n)$ are then modified in a gradient descent algorithm to minimize the local error $e_{n+1}$. An illustration of the algorithm is presented in Fig. \ref{fig0}. 
After each iteration of the gradient descent we replace the parameters with local mean values to have a smooth variation of the model parameters with time. In addition, we limit the range of the parameters to $|\Gamma_{ab}(n)|<\Gamma_{max}$ and $|\Lambda_{ab}(n)|<\Lambda_{max}$.

More precisely, we do the following: 
\begin{itemize}
	\item start from an initial set of parameters $\{\Gamma_{ab}(n)=\Lambda_{ab}(n)=0:n=0,\dots,T-1\}$
	
	\item for $t_{GD}$ iterations do:
	
	\begin{enumerate}
		\item for $n=0,\dots,T-1$:
		\begin{itemize}
			\item estimate the local gradients
			
			\begin{align}
			\frac{\partial}{\partial \Gamma_{ab}(n)}e_{n+1} &= -(q_a(n+1)-\tilde{q}_a(n+1))\dot q_b(n) \Delta t,\\
			\frac{\partial}{\partial \Lambda_{ab}(n)}e_{n+1} &= -(q_a(n+1)-\tilde{q}_a(n+1))(q_b(n)-\mu_b(n)) \Delta t,
			\end{align}
			
			\item update the model parameters

			\begin{align}
			\Delta \Gamma_{ab}(n) &=-\eta_n \frac{\partial}{\partial \Gamma_{ab}(n)}e_{n+1},\\
			\Delta \Lambda_{ab}(n) &=-\eta_n \frac{\partial}{\partial \Lambda_{ab}(n)}e_{n+1},
			\end{align}

		\end{itemize}
		
		\item regularization
		\begin{itemize}
			\item smooth the parameters
	
			if $n=0$:
			\begin{align}
			\Gamma_{ab}(n) &\leftarrow [\Gamma_{ab}(n)+\Gamma_{ab}(n+1)]/2,\\
			\Lambda_{ab}(n) &\leftarrow [\Lambda_{ab}(n)+\Lambda_{ab}(n+1)]/2,
			\end{align}
			
			otherwise:
			\begin{align}
			\Gamma_{ab}(n) &\leftarrow [2\Gamma_{ab}(n)+\Gamma_{ab}(n-1)+\Gamma_{ab}(n+1)]/4,\\
			\Lambda_{ab}(n) &\leftarrow [2\Lambda_{ab}(n)+\Lambda_{ab}(n-1)+\Lambda_{ab}(n+1)]/4,
			\end{align}
	
			\item limit the parameters
	
			\begin{align}
			\Gamma_{ab}(n) &\in (-\Gamma_{max},+\Gamma_{max}),\\
			\Lambda_{ab}(n) &\in (-\Lambda_{max},+\Lambda_{max}).
			\end{align}

\end{itemize}

	\end{enumerate}
	
\end{itemize}

The learning rate $\eta_n$ is a positive number which can slowly increase with $n$ as magnitude of the gradients approaches to zero.
We can follow the value of error function $E$ to be sure that it is decreasing with the number of GD iterations. We stop the learning process in case the error function is growing and take the optimal parameters where the error function is minimum. The time complexity of the learning algorithm is proportional to $t_{GD}TN$ for sparse interaction graphs $\Lambda$ and $\Gamma$.
 
In the next section we apply the above algorithm to estimate the model parameters which are best to describe the synthetic and empirical data by a Lagrangian dynamics. In addition, the dynamical model is used to compute the response of local quantities $q_a(n)$ and $\dot q_a(n)$ to changes in the initial values $q_b(0)$ and $\dot q_b(0)$.

\begin{figure}
	\includegraphics[width=14cm]{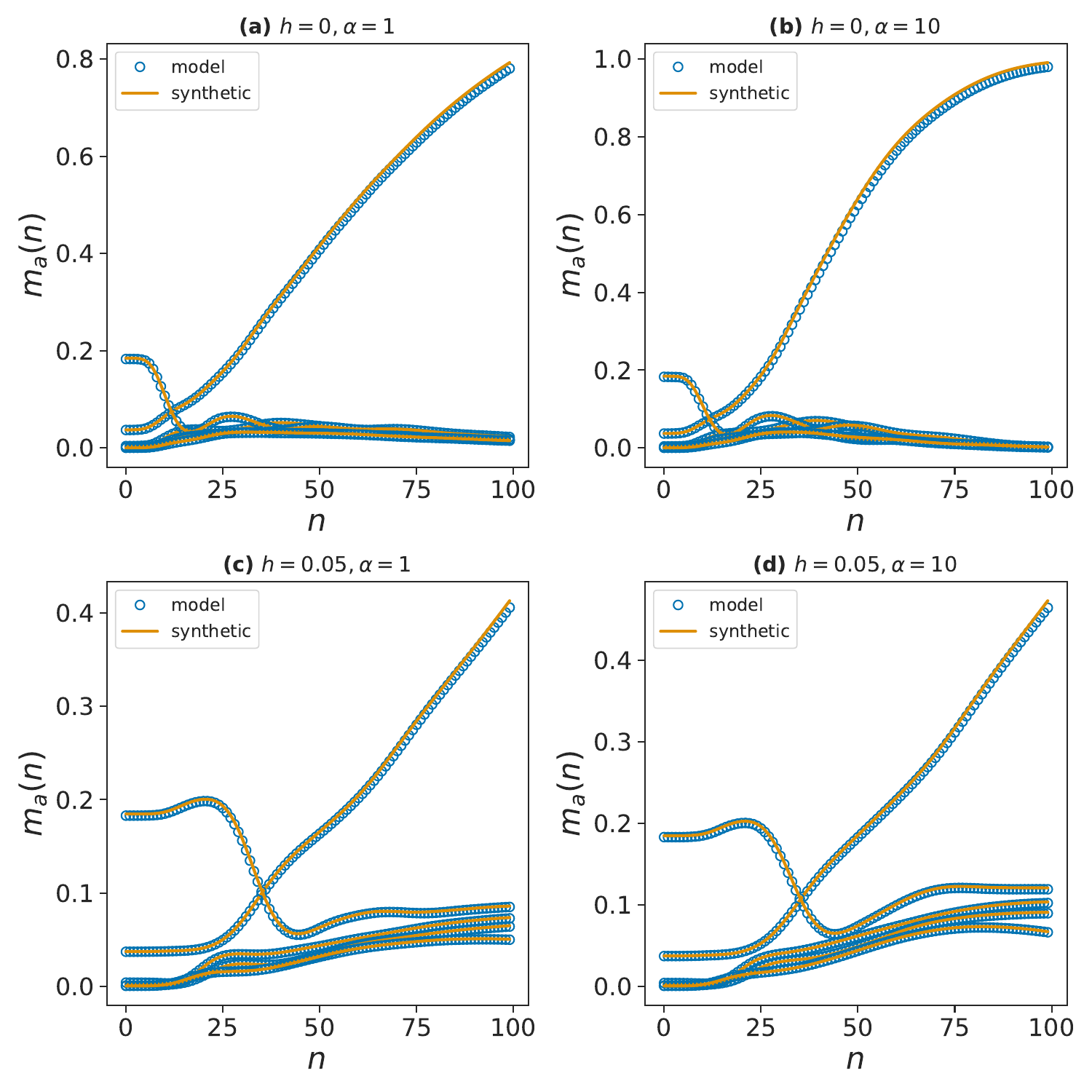} 
	\caption{Comparing the Lagrangian (model) and the synthetic dynamics in a movement process of $T=100$ time steps. The linear size of lattice is $L=10$ and total population is $M=10^4$. Local population densities $m_a(n)$ are plotted vs times step $n$ for a few sites around the lattice center $x_D=y_D=L/2$ and different parameters $h,\alpha$ in the synthetic dynamics.}\label{fig1}
\end{figure}

\begin{figure}
	\includegraphics[width=16cm]{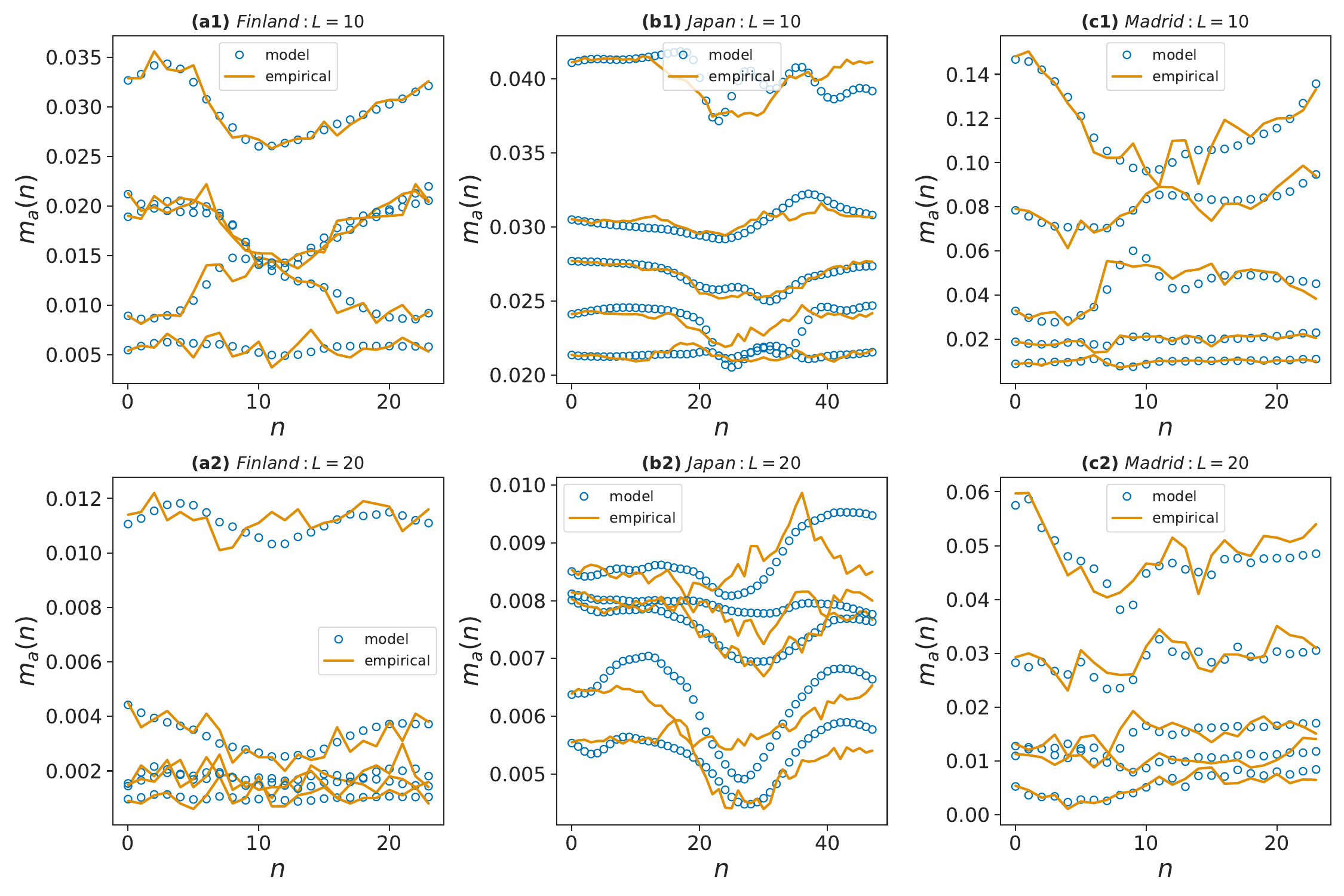} 
	\caption{Comparing the Lagrangian (model) and the empirical dynamics in movement processes of $T=48$ (Japan) and $T=24$ (Finland, Madrid) time steps. (upper panels) The linear size of lattice is $L=10$. (lower panels) The linear size of lattice is $L=20$. Local population densities $m_a(n)$ are plotted vs times step $n$ for a few sites around the lattice center $x_D=y_D=L/2$.}\label{fig1-ma}
\end{figure}

\begin{figure}
	\includegraphics[width=14cm]{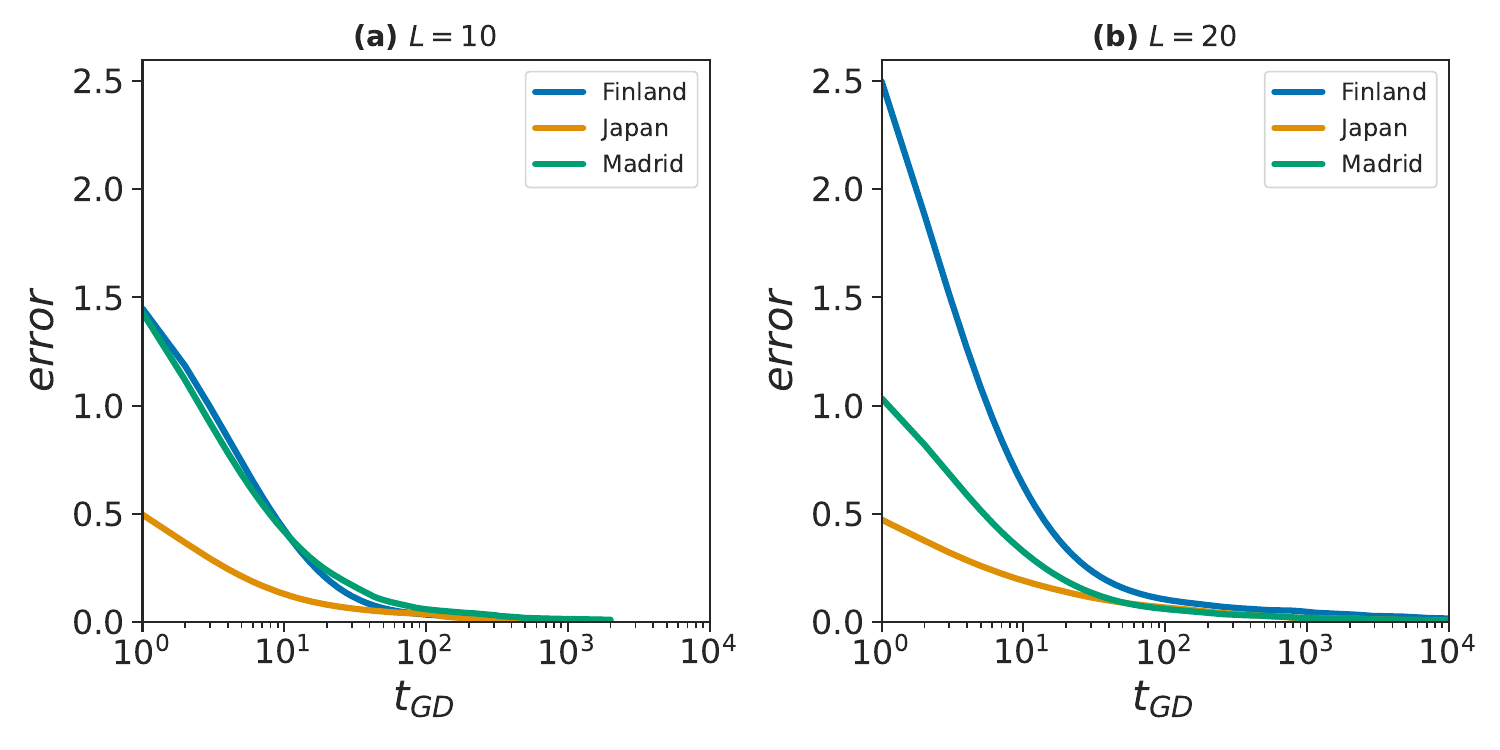} 
	\caption{Error function $E$ vs the number of gradient descent iterations $t_{GD}$ in modeling of the empirical data. (a) The linear size of lattice is $L=10$. (b) The linear size of lattice is $L=20$.}\label{fig1-en}
\end{figure}

\section{Results}\label{S4}
Let us start with modeling a synthetic movement process of $T=100$ time steps as described in Sec.\ref{S1}. All numerical simulations of the synthetic dynamics are performed with total population $M=10^4$ on a lattice of linear size $L=10$. 
Figure \ref{fig1} shows how the Lagrangian model and the inference algorithm reproduce a single realization of the synthetic dynamics. Here we run the learning algorithm for $t_{GD}=2000$ iterations with an increasing learning rate $\eta_n=0.001+0.0015 n$. The best value of the learning rate in general depends on the dynamics and is obtained by error and trial in order to minimize the error function. We run the algorithm for a few times and keep the parameters which result in the smallest error during the learning process. The model parameters are limited by $\Gamma_{max}=\Lambda_{max}=N$. The figure displays time evolution of population density for a few sites around the center when control parameters $\alpha$ and $h$ are varied. Recall that for larger values of $\alpha$ the agents are more likely to choose a neighboring site that is closer to the destination. And, increasing $h$ enhances the strength of interactions and so the waiting times, depending on the present load of the links.   

In Figs. \ref{fig1-ma} and \ref{fig1-en} we report the local populations $m_a$ and the error values obtained by the Lagrangian dynamics for a single realization of the empirical data with different lattice sizes $L=10, 20$. Higher errors are observed when the relative variations in the local populations are considerable. The method is better at describing a mesoscopic dynamics with smooth variation of population in space and time. The error function decreases to values smaller than $10^{-2}$ with the number of iterations $t_{GD}$ as the learning rate $\eta_n$ increases linearly to $\simeq 1-2$.

\begin{figure}
	\includegraphics[width=14cm]{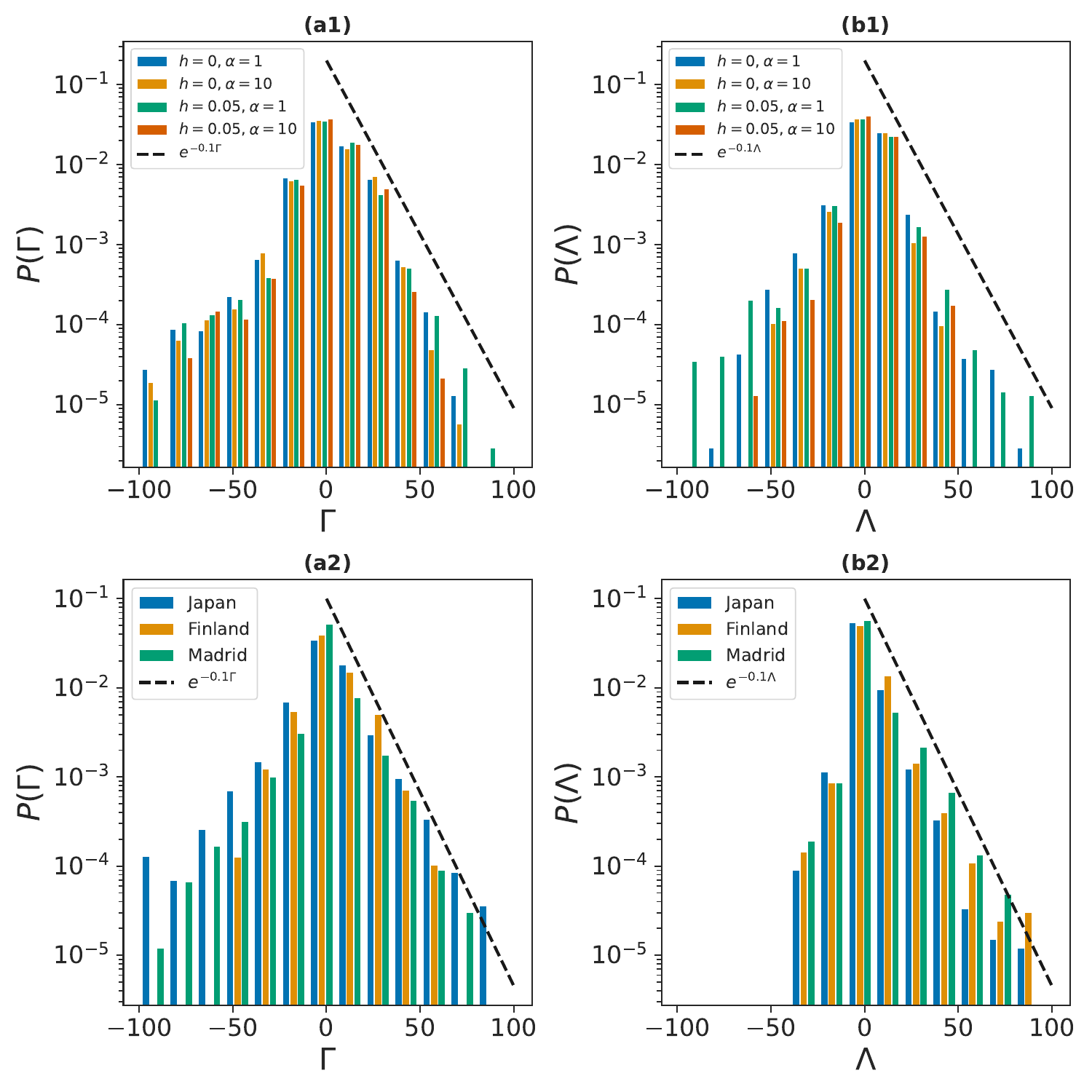} 
	\caption{Probability distribution of the model parameters ($\Gamma_{ab}$ and $\Lambda_{ab}$) with lattices of linear size $L=10$. ((a1),(b1)) From modeling of the synthetic dynamics. ((a2),(b2)) From modeling of the empirical data. Dashed lines display exponential functions for reference to compare with the probability distributions.}\label{fig2}
\end{figure}

\begin{figure}
	\includegraphics[width=14cm]{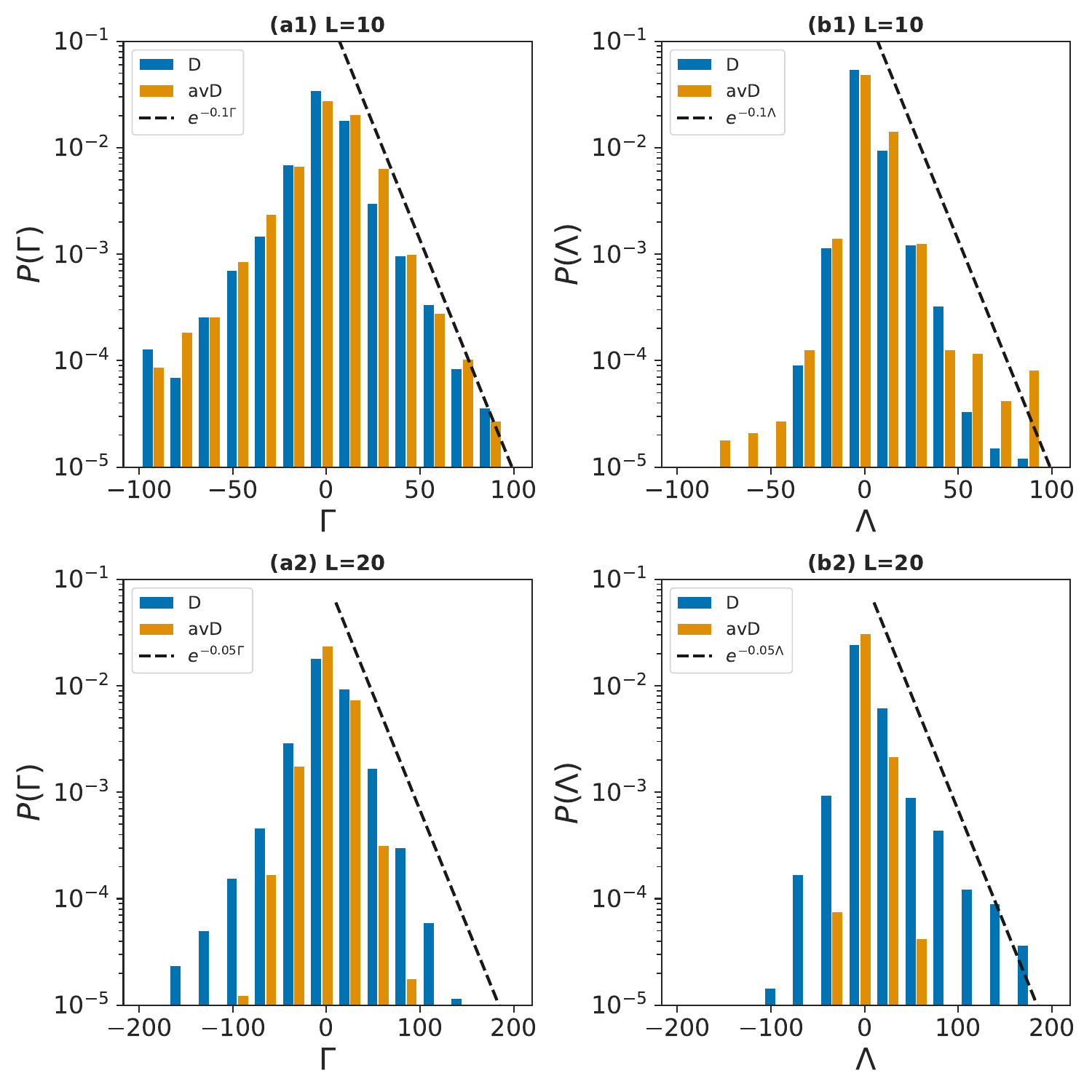} 
	\caption{Probability distribution of the model parameters ($\Gamma_{ab}$ and $\Lambda_{ab}$) from modeling of the empirical data from Japan. The lattices have linear size $L=10$ ((a1),(b1)) and $L=20$ ((a2),(b2)). The distributions are obtained for a single day (D) and the average dynamics over $75$ working days (avD). Dashed lines display exponential functions for reference to compare with the probability distributions.}\label{fig2-J}
\end{figure}

Distribution of the inferred model parameters $\Gamma_{ab}, \Lambda_{ab}$ for a single realization of the synthetic and empirical data are reported in Fig. \ref{fig2}. The lattice size here is $L=10$. All distributions decay exponentially from the maximum value at zero magnitude for the parameters. The rate of exponential decay is well around $0.1$ for all the reported cases, identifying a characteristic range of parameters $\Lambda^*, \Gamma^* \simeq 10$. Notably, the real data from Japan exhibits a distribution of interaction parameters $P(\Lambda)$ which is more concentrated on zero values in contrast to that of Finland and Madrid.

Note that we are using the same lattice size $L$, and so the same number of parameters ($\Gamma_{ab}, \Lambda_{ab}$), to model the empirical data which have different spatial resolutions for Finland, Japan, and Madrid. Moreover the model that describes a single realization of the dynamical process could be different from the one that is inferred from the average dynamics. Figure \ref{fig2-J} shows that the above picture holds qualitatively also for larger lattice size $L=20$ and the average dynamics (over $75$ working days) of the empirical data from Japan. Note that the characteristic range of parameters increases with $L$ as expected; larger parameters are needed to describe the population variations in smaller spatial scales.  

\begin{figure}
	\includegraphics[width=14cm]{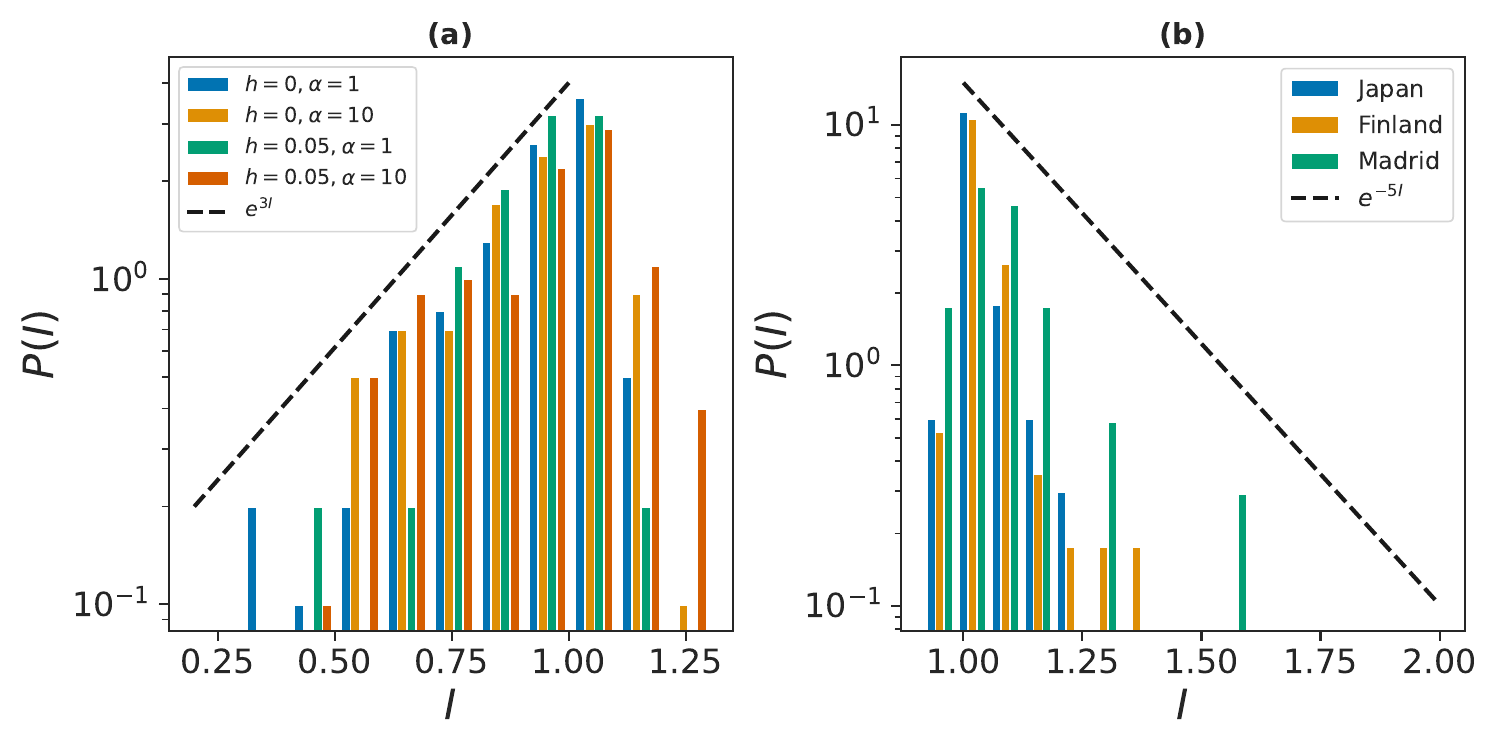} 
	\caption{Probability distribution of the local ratio of inertia to dissipation ($I_{a}$) with lattices of linear size $L=10$. (a) From modeling of the synthetic dynamics. (b) From modeling of the empirical data. Dashed lines display exponential functions for reference to compare with the probability distributions.}\label{fig3}
\end{figure}

\begin{figure}
	\includegraphics[width=16cm]{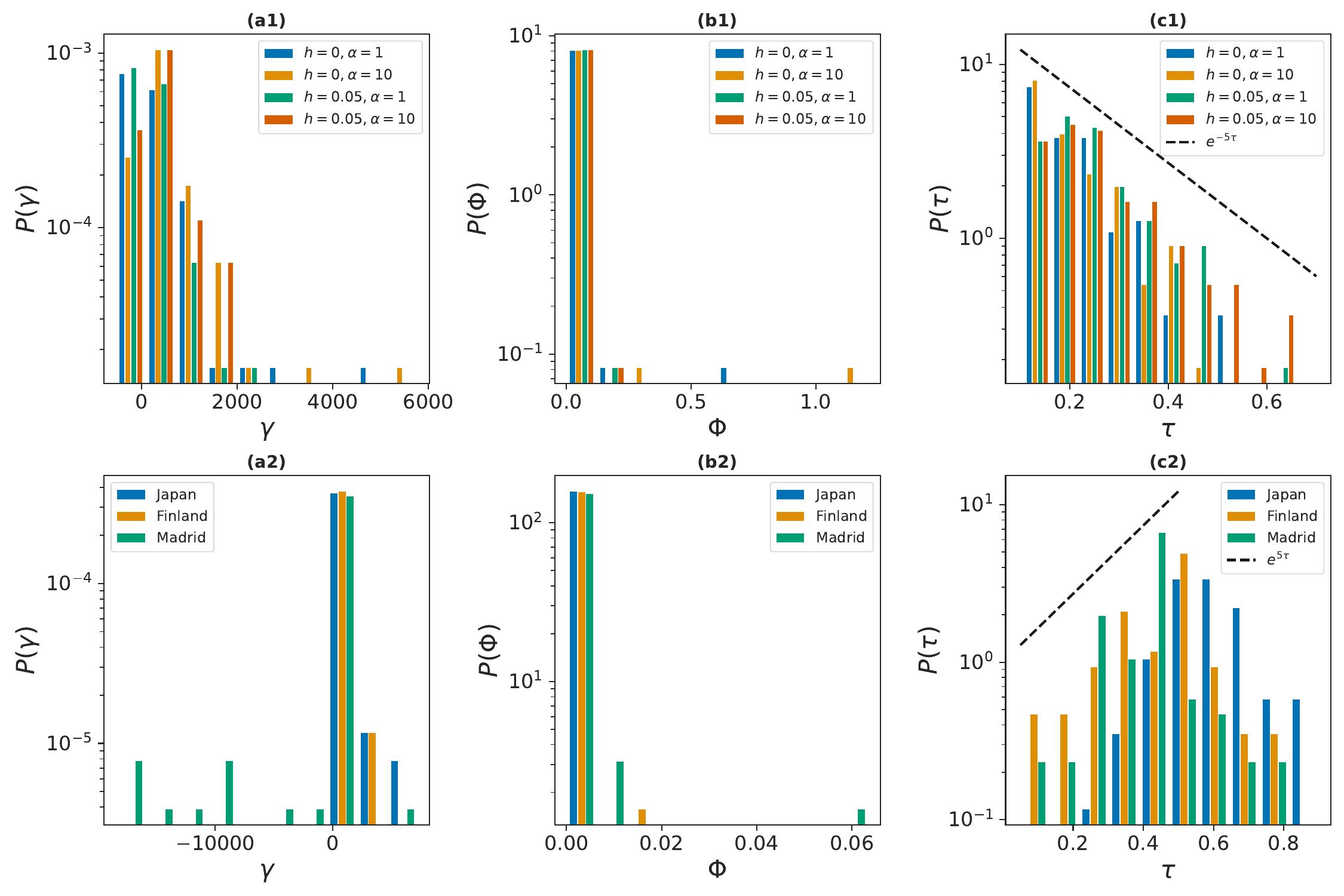} 
	\caption{Probability distribution of the local fluxes ($\Phi_a$), dissipations ($\gamma_a$), and characteristic times ($\tau_a$) with lattices of linear size $L=10$. ((a1),(b1),(c1)) From modeling of the synthetic dynamics. ((a2),(b2),(c2)) From modeling of the empirical data. Dashed lines display exponential functions for reference to compare with the probability distributions.}\label{fig4}
\end{figure}

\begin{figure}
	\includegraphics[width=16cm]{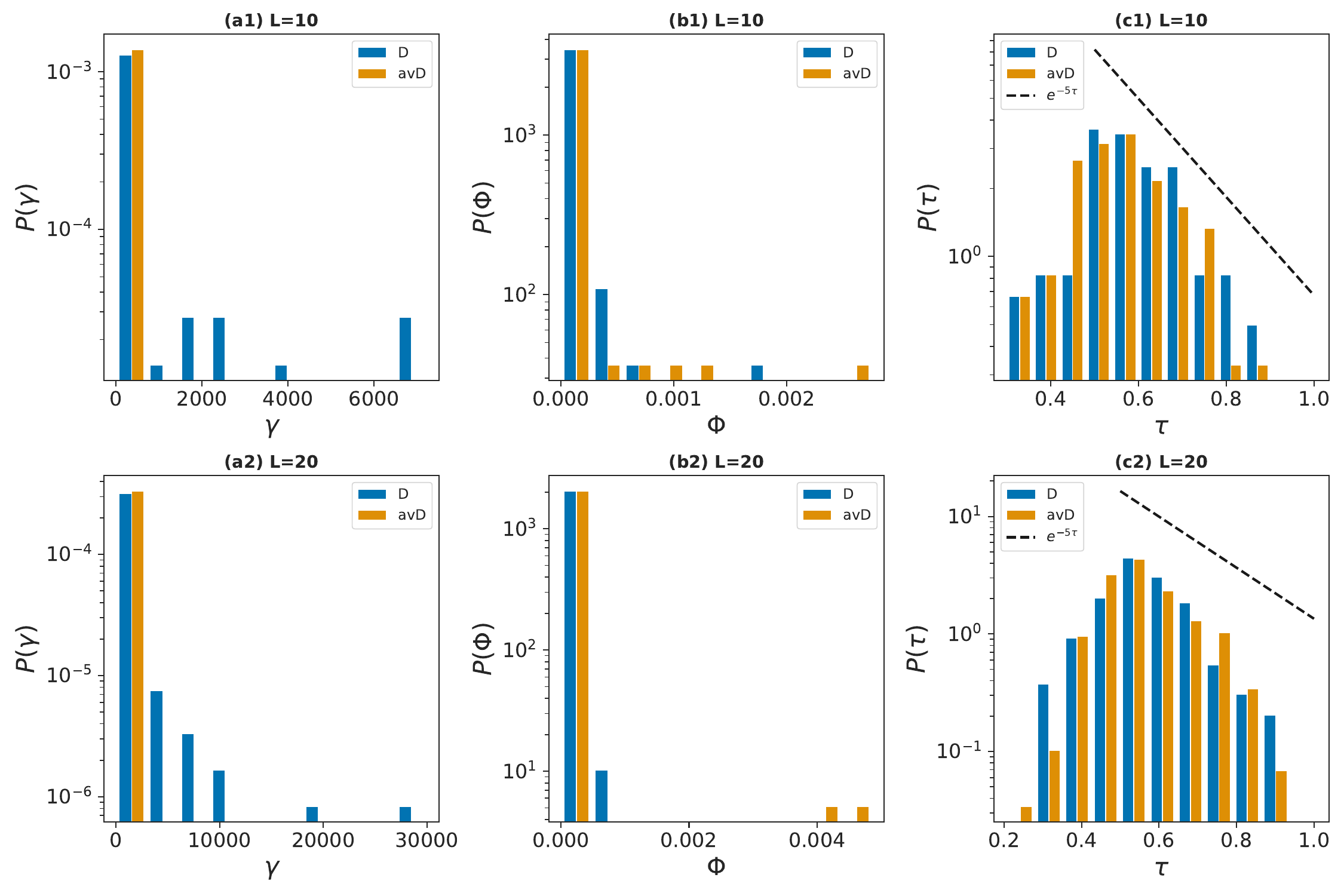} 
	\caption{Probability distribution of the local fluxes ($\Phi_a$), dissipations ($\gamma_a$), and characteristic times ($\tau_a$) from modeling of the empirical data (Japan). The lattices have linear size $L=10$ ((a1),(b1),(c1)) and $L=20$ ((a2),(b2),(c2)). We show the results for a single day (D) and the average dynamics over $75$ working days (avD). Dashed lines display exponential functions for reference to compare with the probability distributions.}\label{fig4-J}
\end{figure}

To check the importance of inertia in the dynamics we study the ratios
\begin{align}
	I_a =  \frac{\int_0^1 dt| \ddot{q}_a|}{\int_0^1 dt |\sum_{b\in \{a,\partial a\}}\Gamma_{ab}\dot q_b|}.
\end{align}
That is the time average of magnitudes of local acceleration and dissipation in the equations of motion.
Figure \ref{fig3} displays the distribution $P(I)$ of the above quantity for a single realization of the synthetic and empirical dynamical data with lattice size $L=10$. We observe that in all cases inertia and dissipation are of the same order and we can not ignore the role of inertia in the dynamics. The presence of interactions and selection of shortest routes at the same time results in larger values of accelerations compared to dissipation in the synthetic data. Moreover, the smaller variance of the $I_a$ in Japan separates it from the larger variances in Finland and Madrid.

The product of velocity $\dot q_a$ and dissipative forces $\Gamma_{ab}\dot q_b$ provides a local measure of dissipation $\gamma_a$ at site $a$. Time averaging in the process yields 
\begin{align}
	\gamma_a = \sum_{b\in \{a,\partial a\}} \int_0^1 \dot q_a(t)\Gamma_{ab}(t)\dot q_b(t) dt.
\end{align}
The fraction of sites in the lattice with a given level of dissipation $\gamma$ is denoted by $P(\gamma)$.
High velocities around site $a$ result in a large value of local dissipation when $\Gamma_{ab}$ is considerable. Looking at the distributions of $\gamma_a$ and $\Gamma_{ab}$ helps to identify such correlated regions in the system.  

The flow at site $a$ is obtained by product of the velocity $\dot m_a$ and the change $dm_a$ in a small time interval $dt$. The time averaged flow reads
\begin{align}
	\Phi_a=\int \dot m_a(t) dm_a=\int_0^1 (\dot m_a(t))^2 dt.
\end{align}
A large value of $\Phi_a$ is obtained for high velocities at site $a$ in the dynamical process. Note that the velocity of neighbors and the parameters $\Gamma_{ab}$ are not involved in definition of the average flow $\Phi_a$. 

It is also important to know the temporal ordering of the main flows in the system, besides the magnitudes $\Phi_a$. Time dependence of the local flows can be used to define the characteristic times
\begin{align}
	\tau_a = \frac{\int_0^1 t(\dot m_a(t))^2 dt}{\int_0^1 (\dot m_a(t))^2 dt}.
\end{align}
Here $\tau_a$ provides an estimate of the typical time of the largest flow at site $a$; uniform flows in the time interval of the process $t\in (0,1)$ give $\tau_a=1/2$. 
  
Figure \ref{fig4} shows the above quantities in the Lagrangian models inferred from a single realization of the synthetic and empirical data with lattice size $L=10$. Besides the very large and positive dissipations, we also observe negative values specially in Madrid. The flows are concentrated around zero except for a few sites which are near the center. As expected, the two distributions $P(\gamma)$ and $P(\Phi)$ show that large values of $\gamma$ are accompanied by large flows $\Phi$. In panels (a1) and (b1) we see that optimality of the movements (large $\alpha$) in the synthetic dynamics shifts the local dissipation $\gamma$ to higher positive values whereas the flows $\Phi$ are mostly concentrated around zero. In words, such a movement is giving rise to high velocities for neighboring sites $a,b$ with a large coupling $\Gamma_{ab}$.       

For the synthetic data, which come from movements toward the center, the relevant flows are mostly observed at the beginning of the movement process (small $\tau$). On the other hand, for the empirical data that span a whole day, the characteristic times are closer to the middle of the process ($\tau=1/2$). Nevertheless in all cases $P(\tau)$ displays an exponential behavior with a decay rate that is around $5$ as shown in figure \ref{fig4} (panels (c1),(c2)). This defines the time scale $\tau^*\simeq 0.2$ for significant flows in the system. Here again Japan can easily be distinguished from Finland and Madrid by displaying a larger dissipation, smaller flows, and larger time scales.

To see how the spatial resolution affects the above distributions, in Fig. \ref{fig4-J} we report the result for two lattice sizes $L=10, 20$ which are obtained by the empirical data from Japan. We also compare the results for a single day with that of the average dynamics (over $75$ working days) to check the role of statistical fluctuations in a single day. We observe that larger values of dissipation $\gamma$ and flow $\Phi$ appear as $L$ increases and $P(\tau)$ is more concentrated around $\tau=1/2$ for the larger lattice size. On the other hand, the average dynamics displays a smaller variability in $\gamma$ but still a broad range of $\Phi$ and $\tau$ values are observed as for the case of a single day.

\subsection{Dynamical susceptibilities}\label{S41}
\begin{figure}
	\includegraphics[width=14cm]{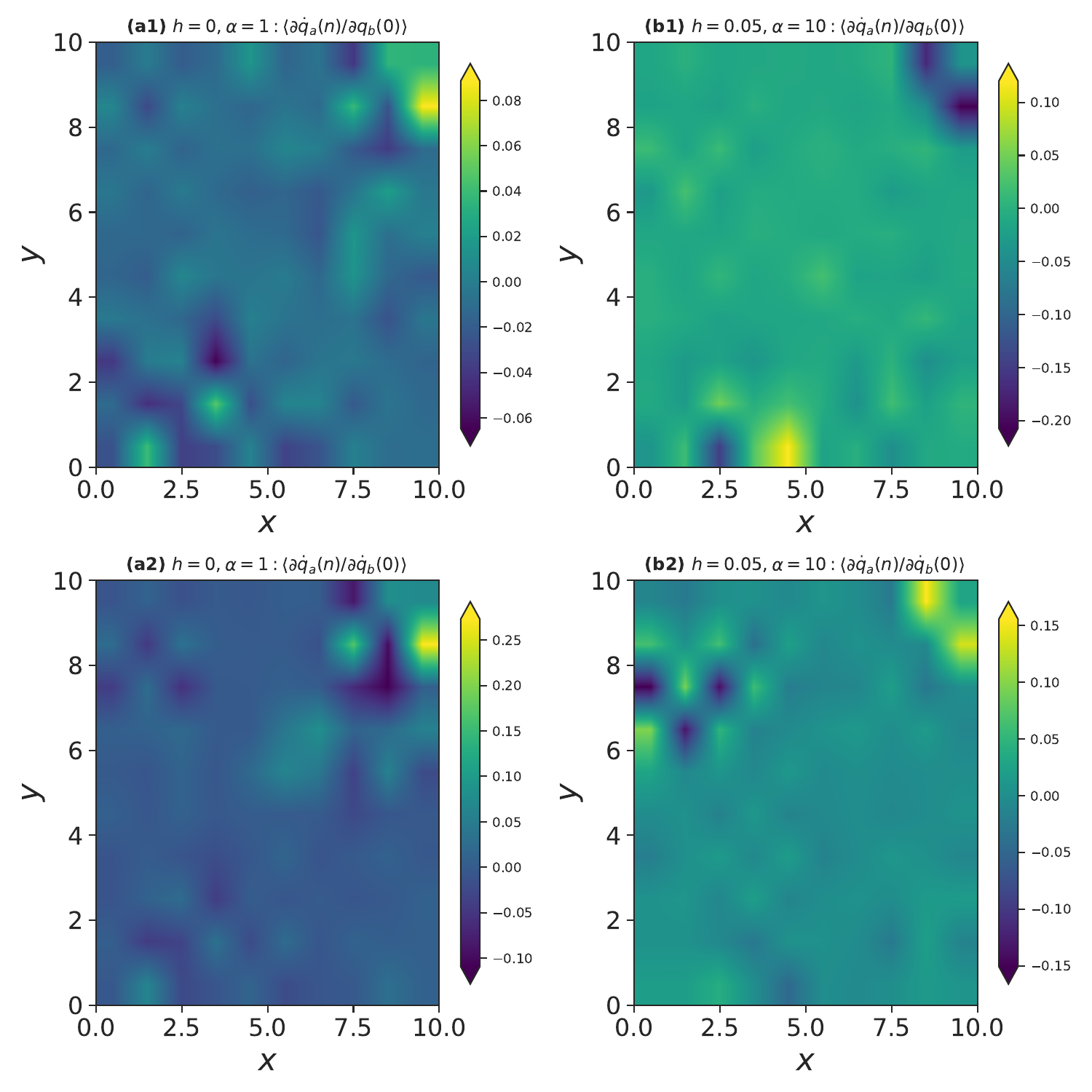} 
	\caption{Color map of the average response of velocities $\dot q_a(n)$ to initial populations ($q_b(0)$) and velocities ($\dot q_b(0)$) with lattices of linear size $L=10$. From modeling of the synthetic dynamics with parameters:  ((a1),(a2)) $h=0, \alpha=1$ and  ((b1),(b2)) $h=0.05, \alpha=10$. Variables $x,y \in [0,L-1]$ show position of sites in the lattice.}\label{fig5}
\end{figure}

The dynamical equations (\ref{EC1},\ref{EC2}) can be used to estimate the response of local populations to changes in the initial values of the movement process. Such information would be useful in providing a measure of dynamical sensitivity and criticality in a movement process. Consider for instance the case of variations in the initial velocities $\dot q_a(0)$ and define 
\begin{align}
\chi_{ab}(n) &=\frac{\partial q_a(n)}{\partial \dot q_b(0)},\\
\psi_{ab}(n) &=\frac{\partial \dot q_a(n)}{\partial \dot q_b(0)}.
\end{align}
From the dynamical equations, we obtain the following recursive relations between the susceptibilities 
\begin{align}\label{dEC} 
	\psi_{ac}(n+1) &=\psi_{ac}(n)- \sum_{b \in \{a,\partial a\}}\Gamma_{ab}(n)\psi_{bc}(n) \Delta t
	- \sum_{b \in \{a,\partial a\}}\Lambda_{ab}(n) \chi_{bc}(n)\Delta t,\\
	\chi_{ab}(n+1) &=\chi_{ab}(n)+\psi_{ab}(n+1)\Delta t.
\end{align}
The equations are solved for the susceptibilities in a sequential way, starting from the initial condition 
\begin{align}
	\chi_{ab}(0) &=0,\\
	\psi_{ab}(0) &=\delta_{a,b}.
\end{align}
In this way we obtain the whole set of susceptibilities for different time steps $n$ given a single realization of the dynamical process. The time complexity of this computation is of order $TN^2$ for sparse interaction matrices $\Lambda, \Gamma$.  
The time average of local susceptibilities are then defined as follows
\begin{align}
	\chi_{a} &=  \frac{1}{NT}\sum_b \sum_n \chi_{ab}(n)=\langle \frac{\partial q_a(n)}{\partial \dot q_b(0)}\rangle,\\
	\psi_{a} &=  \frac{1}{NT}\sum_b \sum_n \psi_{ab}(n)=\langle \frac{\partial \dot q_a(n)}{\partial \dot q_b(0)}\rangle.
\end{align}
Figure \ref{fig5} displays the average response of velocities $\dot q_a$ to small changes in $q_b(0)$ and $\dot q_b(0)$ for a single realization of the synthetic dynamics with lattice size $L=10$. We observe positive and negative values of local susceptibilities with magnitudes of order $0.1$. We see that highly susceptible regions (positive or negative) are usually clustered and the two average susceptibilities $\langle \frac{\partial \dot q_a(n)}{\partial q_b(0)}\rangle$ and $\langle\frac{\partial \dot q_a(n)}{\partial \dot q_b(0)}\rangle$ are strongly correlated. Similar behaviors are observed also for the average response of local populations $q_a$ to $q_b(0)$ and $\dot q_b(0)$ (not shown here).

\begin{figure}
	\includegraphics[width=16cm]{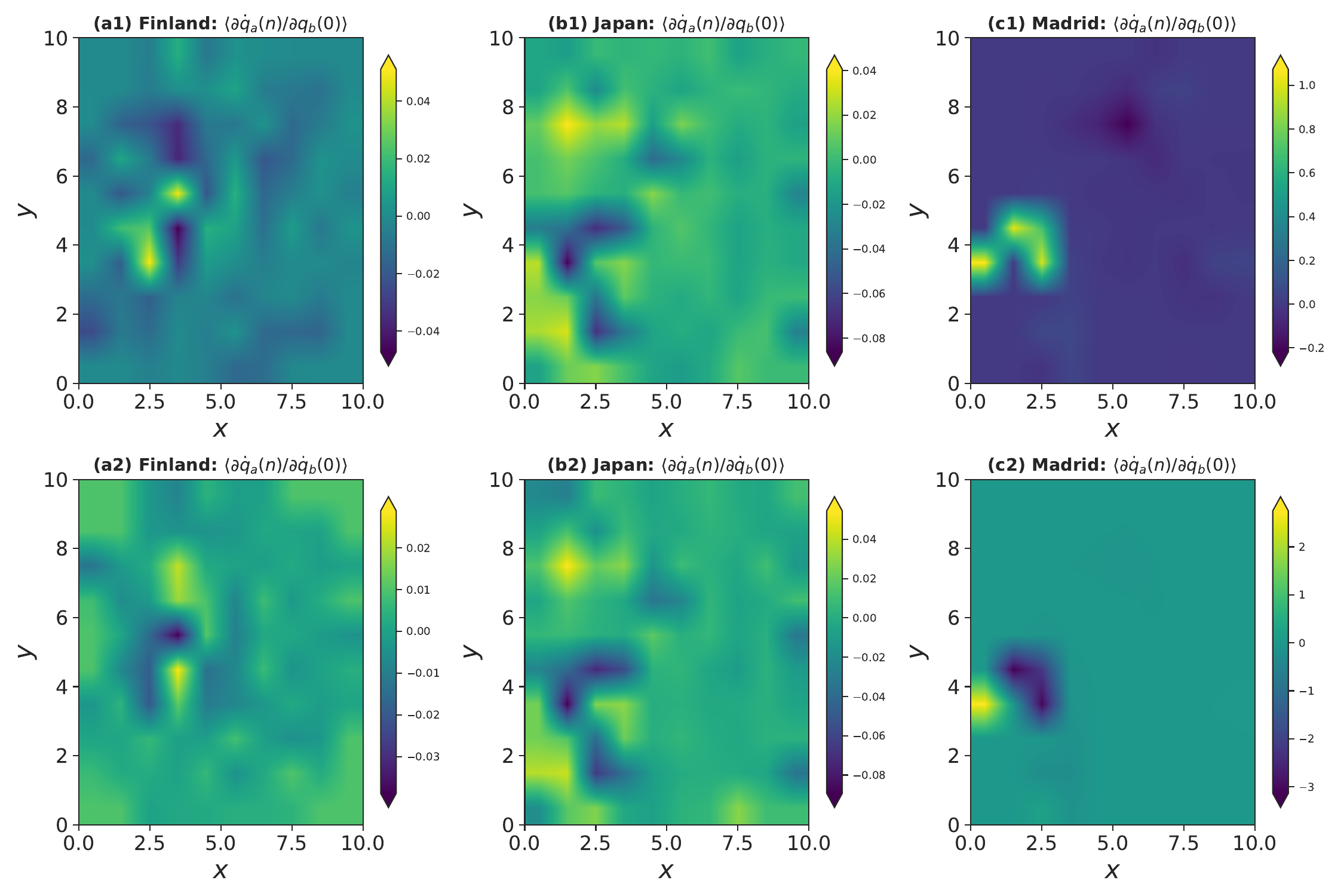} 
	\caption{Color map of the average response of velocities $\dot q_a(n)$ to initial populations ($q_b(0)$) and velocities ($\dot q_b(0)$) with lattices of linear size $L=10$. From modeling of the empirical data: ((a1),(a2)) Finland, ((b1),(b2)) Japan, and ((c1),(c2)) Madrid. Variables $x,y \in [0,L-1]$ show position of sites in the lattice.}\label{fig6-L10}
\end{figure}

\begin{figure}
	\includegraphics[width=16cm]{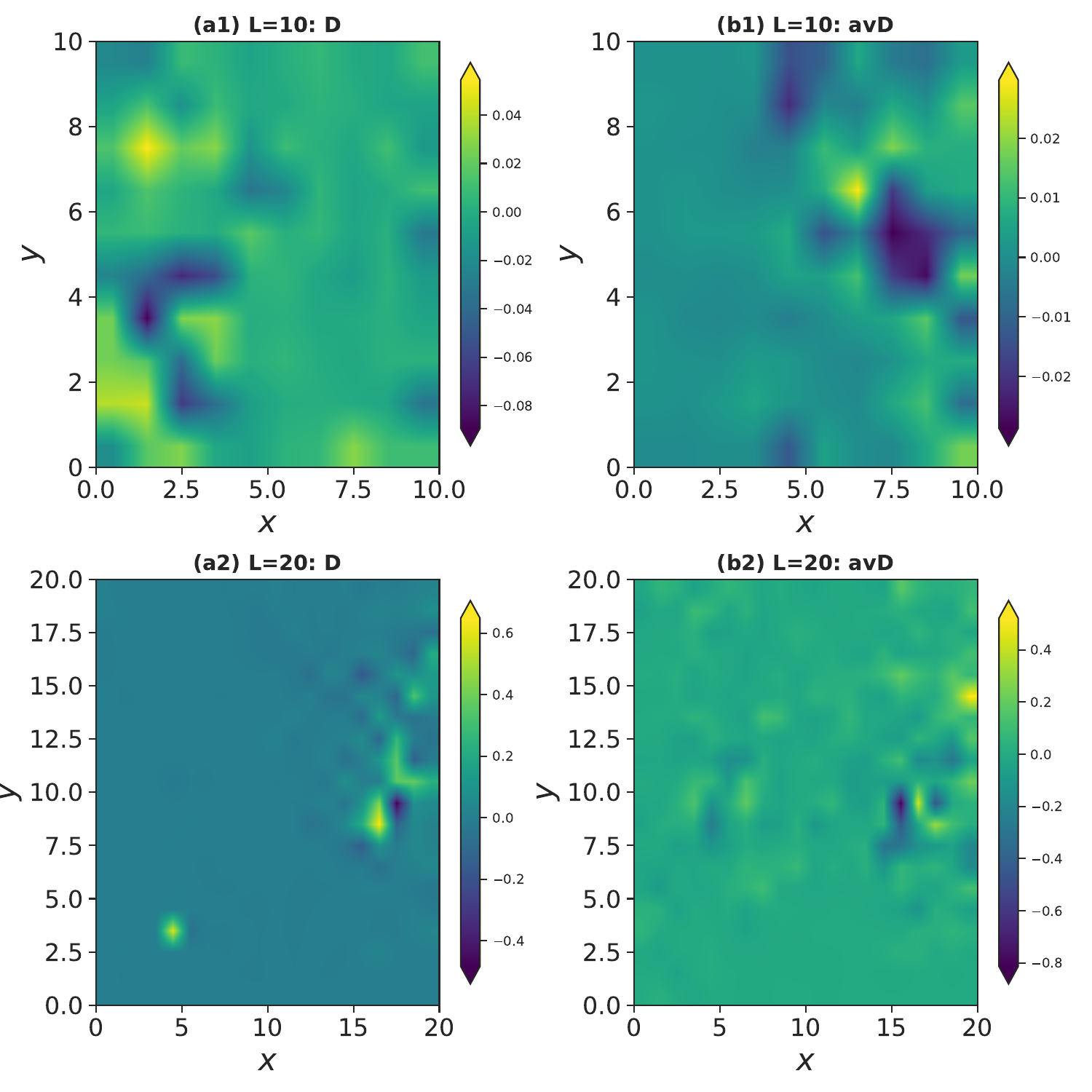} 
	\caption{Color map of the average response of velocities $\dot q_a(n)$ to initial ones ($\dot q_b(0)$) from modeling of the empirical data (Japan). The lattices have linear size $L=10$ ((a1),(b1)) and $L=20$ ((a2),(b2)). Panels ((a1),(a2)) show the results for a single day (D). Panels ((b1),(b2)) show the results for the average dynamics over $75$ working days (avD). Variables $x,y \in [0,L-1]$ show position of sites in the lattice.}\label{fig6-J}
\end{figure}

In Fig. \ref{fig6-L10} we report the susceptibilities for a single realization of the empirical dynamics with lattice size $L=10$. The magnitude of responses in Madrid are much larger than those of Finland and Japan, probably because of the smaller resolution of the Madrid data. Figure \ref{fig6-J} shows how the lattice size and averaging over different dynamical realizations in Japan affect the susceptibilities. As expected increasing the resolution by $L$ results in higher responses. On the other hand, averaging lead to a smoother dynamics and therefore smaller responses. This is more pronounced for size $L=10$ compared to $L=20$.

\section{Conclusion}\label{S5}
In summary we developed an effective Lagrangian formalism to explain time variations of population distribution in a movement process in terms of an interpretable potential and dissipation function. A dynamical gradient descent algorithm was used to estimate the model parameters and anticipate the system susceptibilities to local changes in the initial values of the movement process. Modeling of synthetic and empirical data revealed the significance of inertia in describing these movements and showed how microscopic interactions and route optimization shape the mesoscopic dynamics. 

Having an effective field theory that can efficiently reproduces the coarse-grained dynamics of a movement process would be helpful to study the large-scale performances of such dynamical processes; for instance by looking at variations of the associated Hamiltonian as the model parameters and the state of system change with time. It would be interesting to see how deviations from the stationary equations of motion affect the efficiency and entropy production of a movement process \cite{BR-srep-2020,BNR-jphys-2021,NR-pre-2025}. Moreover, the statistical behavior of dynamical susceptibilities would be helpful to say something about the criticality or complexity of a movement process. 
         
We used the Euler-Cromer method to find an approximate solution to the equation of motions. Numerical methods like the fourth-order Runge-Kutta can provide a more accurate approximation of the dynamics. Finally, note that the above formalism can be used to model  time variations of any probability distribution by a Lagrangian dynamics.

\acknowledgments
This work was performed using the ALICE compute resources provided by Leiden University.

\appendix
%\counterwithin{figure}{section}

\section{The synthetic dynamics}\label{app-A}

We consider a two-dimensional square lattice of linear size $L$ with $N=L\times L$ nodes indexed by $a=1,\dots,N$ or coordinates $x_a,y_a \in [0,L-1]$. The set of neighbors of node $a$ is denoted by $\partial a$ with node degree $k_a=|\partial a|$. We use the growth model of Ref. \cite{Stanley-natc-2017} to generate an initial population distribution $M_a(0)$, which closely resembles the empirical distributions. The model starts with a unit of population at the center of network $D$ with coordinates $x_D=y_D=L/2$, that is $M_D=1$ and $M_{a\neq D}=0$. Then, in each step of the growth algorithm a unit of population is added to a randomly chosen site $a$ with probability $\propto (M_a+C)$, if $M_b>0$ for a site $b$ with $|x_b-x_a|\le R$ and $|y_b-y_a|\le R$. Here we take the parameters $C=0.5$ and $R=1$. Different values of $C$ change the exponent of the average population density at distance $r$ from the center  $\rho(r)\propto r^{-\beta}$ as define in Ref. \cite{Stanley-natc-2017}. This is a power law distribution which is not very sensitive to value of $R$. For $C=0.5$ one obtains $\beta \simeq 0.2$ which is around the empirical values for London ($\beta\simeq 0.3$) and Beijing ($\beta \simeq 0.1$). The algorithm continues until the total population is $M=\sum_a M_a$. The $M$ units of population or agents are indexed by $i=1,\dots,M$.

As a reference dynamics we consider a movement process of $T$ time steps where each agent (driver) moves toward a single destination $D$ which here is the center of network at $x_D=y_D=L/2$. The movement process starts with the initial distribution of drivers $M_a(0)$ which is obtained by the above growth model. We use the dynamical model of Ref. \cite{Saad-prr-2020} to move the agents according to the distances of neighboring sites to the destination. More precisely, the probability of choosing a neighbor $b$ of site $a$ is 
\begin{align}
	p_{a\to b}=\frac{e^{-\alpha (D_b-D_a)}}{ \sum_{c\in \partial a} e^{-\alpha (D_c-D_a)}},
\end{align}
where $D_a=|x_D-x_a|+|y_D-y_a|$ is the Manhattan distance of node $a$ from the destination $D$. The parameter $\alpha\ge 0$ controls the degree of closeness to the destination. 

The waiting time $\Delta$ that the driver spends in link $(ab)$ is drawn from a Poisson distribution 
\begin{align}
	P_{ab}(\Delta|\rho_{ab}(n))= e^{-\tau_{ab}}\frac{\tau_{ab}^{\Delta}}{\Delta!}.
\end{align}
The mean value $\tau_{ab}=L+h\rho_{ab}(n)$ depends on the average load $\rho_{ab}(n)$ of link $(ab)$ at time step $n$. The parameter $h\ge 0$ controls the strength of interactions in this system.

The average flux of drivers which exit link $(ab)$ at time step $n$ and arrive at site $b$ is denoted by $f_{ab}(n)$. The center is a sink receiving only incoming fluxes. The dynamical equations governing the loads and fluxes read as follows
\begin{align}
	\rho_{ab}(n) &= p_{a\to b} \sum_{c\in\partial_a,c\neq D} f_{ca}(n-1)+(\rho_{ab}(n-1)-f_{ab}(n-1)),\\
	f_{ab}(n) &=\sum_{n'=1}^n [\rho_{ab}(n')-(\rho_{ab}(n'-1)-f_{ab}(n'-1))]P_{ab}(n-n'|\rho_{ab}(n')) + \rho_{ab}(0)P_{ab}(n|\rho_{ab}^0).
\end{align}
The initial values at time step $n=0$ are given by $f_{ab}(0)=0$ and $\rho_{ab}(0)=(1-\delta_{a,D})M_a(0)/k_a$, where $\delta_{a,b}=1$ if $a=b$, otherwise it is zero. The local population at each time step then is given by $M_a(n)=\sum_{b\in \partial a}\rho_{ab}(n)$ when $a\ne D$. For the destination we have $M_D(n)=M_D(n-1)+\sum_{a\in \partial D}f_{aD}(n)$.

\section{The empirical dynamics}\label{app-B}
Three sets of human mobility data are used in this study to model with a Lagrangian dynamics. 

(i) Data set from Japan \cite{data-Japan}: The dataset provides 75 days of continuous trajectories, with a spatial resolution of 500 × 500 meter grid cells ($200 \times 200$ lattice) and a temporal resolution of 30-minute timeslots (48 per day). We take the dynamical data for all working days from data file "yjmob100k-dataset1.csv".

(ii) Data set from Finland \cite{data-Finland}: The dataset provides temporally dynamic population distribution data for the Helsinki Metropolitan Area at the resolution of $250 \times250$ meter grid cells ($130 \times 101$ lattice). It includes three daily cycles: regular workdays (Mon-Thu), Saturdays, and Sundays. Each cycle has a full 24-hour profile, discretized into one-hour intervals (H0-H23). Each field represents the proportional distribution of the total population across all grid cells for that hour. We take the dynamical data for a single day from data file "HMA \_ Dynamic \_ population \_ 24H \_ workdays.csv".

(iii) Data set from Madrid \cite{data-Madrid}: The Madrid Traffic Dataset (MTD) covers the period from June 1, 2022, to February 29, 2024. It integrates different sources including traffic sensors, meteorological observations, calendar data, road infrastructure, and geographical data. The data are given by nearly $30 \times 30$ meter grid cells ($24 \times 24$ lattice). We take the dynamical data for a single day from data file "MTD \_ complete \_ data.csv".

\end{document}